\documentclass[a4paper,11pt]{article}
\usepackage{aaskaiid2}
\usepackage{xcolor}
\usepackage{enumitem}
\usepackage{bm}
\usepackage{pjournal}
\usepackage{comment}
\usepackage{hyperref}
\usepackage{orcidlink}
\usepackage[utf8]{inputenc}
\DeclareUnicodeCharacter{02BC}{'}

\title{Overview of 21~cm Experiments at high redshift with SKAO}
\ShortTitle{21~cm Experiments at high redshift with SKAO}

\author[]{EoR/CD Science Working Group}
\ShortName{EoR/CD SWG} 
\author[1,2,3]{Gianni Bernardi\orcidlink{0000-0002-0916-7443}}
\author[21, 22]{Daniela Breitman\orcidlink{0000-0002-2349-3341}}
\author[4]{Abhirup Datta\orcidlink{0000-0002-5333-1095}}
\author[5]{Anastasia Fialkov\orcidlink{0000-0002-1369-633}}
\author[7]{L\'{e}on V.E. Koopmans\orcidlink{0000-0003-1840-0312}}
\author[9]{Adrian Liu\orcidlink{0000-0001-6876-0928}}
\author[10]{Yi Mao\orcidlink{0000-0002-1301-3893}}
\author[11]{Garrelt Mellema\orcidlink{0000-0002-2512-6748}}
\author[12,13]{Florent Mertens}
\author[14]{Andrei Mesinger\orcidlink{0000-0003-3374-1772}}
\author[15,16]{Jonathan Pritchard}
\author[17]{Aurel Schneider\orcidlink{0000-0001-7055-8104}}
\author[18,19]{Cathryn M. Trott*\orcidlink{0000-0001-6324-1766}}
\author[20]{Yidong Xu\orcidlink{0000-0003-3224-4125}}
\author[	21	]{	Anshuman	Acharya	\orcidlink{	0000-0003-3401-4884	}}
\author[	22	]{	Satadru	Bag	\orcidlink{	0000-0003-0141-606X	}}
\author[	23	]{	Rennan	Barkana			} 
\author[	24	]{	Nichole	Barry	\orcidlink{	0000-0003-4127-5353	}} 
\author[	58,36	]{	Oliver	Basquette	\orcidlink{	0009-0006-4922-0209	}}
\author[	26	]{	Michele	Bianco	\orcidlink{	0000-0002-6766-0017	}}
\author[	27	]{	Stefanie A.	Brackenhoff	\orcidlink{	0000-0001-7507-6948	}}
\author[	28	]{	Carlo	Burigana	\orcidlink{	0000-0002-3005-5796	}}
\author[	29	]{	Isabella P.	Carucci			} 
\author[	30	]{	Emilio	Ceccotti	\orcidlink{	0000-0002-3351-5778	}}
\author[	31	]{	Arnab	Chakraborty			} 
\author[	32	]{	Samir	Choudhuri			} 
\author[	33	]{	Tirthankar Roy	Choudhury			} 
\author[	34	]{	Benedetta	Ciardi	\orcidlink{	0000-0002-5037-310X	}}
\author[	35	]{	Hector Afonso G.	Cruz	\orcidlink{	0000-0002-1775-3602	}}
\author[	36	]{	Saswata	Dasgupta	\orcidlink{	0000-0001-6461-769X	}}
\author[	37	]{	Kanan K	Datta	\orcidlink{	0000-0002-2238-5146	}}
\author[	39,40,91	]{	Pratika	Dayal	\orcidlink{	0000-0001-8460-1564	}}
\author[	58,36	]{	Eloy	de Lera Acedo	\orcidlink{	0000-0001-8530-6989	}}
\author[	41	]{	Khandakar Md Asif	Elahi	\orcidlink{	0000-0003-1206-8689	}}
\author[	43	]{	Ivelin	Georgiev	\orcidlink{	0000-0002-1950-5039	}}
\author[	44	]{	Sukhdeep Singh	Gill	\orcidlink{	0000-0003-1629-3357	}}
\author[	45	]{	Sambit K.	Giri	\orcidlink{	0000-0002-2560-536X	}}
\author[	46	]{	Ad\'{e}lie	Gorce	\orcidlink{	0000-0002-1712-737X	}}
\author[	71	]{	Quan Guo	\orcidlink{	0000-0003-2858-5090	}}
\author[	47	]{	Caroline	Heneka	\orcidlink{	0000-0001-8883-0583	}}
\author[	48	]{	Ian	Hothi	\orcidlink{	0000-0003-3356-5617	}}
\author[	49	]{	Anne	Hutter	\orcidlink{	0000-0003-3760-461X	}}
\author[	50	]{	Piyanat	Kittiwisit			} 
\author[	51	]{	Bohua	Li	\orcidlink{	0000-0002-3600-0358	}}
\author[	52	]{	Yashrajsinh	Mahida	\orcidlink{	0009-0000-1796-797X	}}
\author[	53	]{	Barun	Maity	\orcidlink{	0000-0002-4682-6970	}}
\author[	54	]{	Suman	Majumdar	\orcidlink{	0000-0001-5948-6920	}}
\author[	55	]{	Avery	Meiksin	\orcidlink{	0000-0002-5451-9057	}}
\author[	56	]{	Romain	Meriot	\orcidlink{	0000-0003-1826-9537	}}
\author[	57	]{	Arnab	Mishra			} 
\author[	58,36	]{	Shikhar	Mittal	\orcidlink{	0000-0002-0247-618X	}}
\author[	59	]{	Rajesh	Mondal 			} 
\author[	60	]{	Lauro	Moscardini	\orcidlink{	0000-0002-3473-6716	}}
\author[	61	]{	Julian	Munoz			} 
\author[	62	]{	Satyapan	Munshi			} 
\author[	63,65	]{	Pravin Kumar	Natwariya	\orcidlink{	0000-0001-9072-8430	}}
\author[	64	]{	Leon	Noble	\orcidlink{	0009-0004-3138-1130	}}
\author[	58,36	]{	Oscar Sage David O'Hara	\orcidlink{	0009-0006-3633-5816	}}
\author[	66	]{	Pierre	Ocvirk	\orcidlink{	0000-0002-8488-504X	}}
\author[	67	]{	Samit Kumar	Pal	\orcidlink{	0000-0002-2271-4165	}}
\author[	68	]{	Yannic	Pietschke	\orcidlink{	0009-0009-0873-7262	}}
\author[	69	]{	Steven G.	Piyanat			} 
\author[	72	]{	Rashmi	Sagar			} 
\author[	73	]{	Rasha M.	Samir	\orcidlink{	0000-0003-2716-8332	}}
\author[	3,74	]{	Mario G.	Santos			} 
\author[	75	]{	Benoit	Semelin			} 
\author[	76	]{	Rahul	Shah	\orcidlink{	0000-0001-7682-9219	}}
\author[	77	]{	Yali	Shao	\orcidlink{	0000-0002-1478-2598	}}
\author[	78	]{	Abinash Kumar	Shaw	\orcidlink{	0000-0002-6123-4383	}}
\author[	79	]{	Hayato	Shimabukuro	\orcidlink{	0000-0003-4850-0656	}}
\author[	58,36	]{	Peter H.	Sims 	\orcidlink{	0000-0002-2871-0413	}}
\author[	81,92,93,94	]{	Tomáš	Šoltinský	\orcidlink{	0000-0001-7703-8929	}}
\author[	82	]{	Guochao	Sun	\orcidlink{	0000-0003-4070-497X	}}
\author[	83	]{	Anshuman	Tripathi	\orcidlink{	0000-0002-5091-9950	}}
\author[	84	]{	Tiziana	Trombetti	\orcidlink{	0000-0001-5166-2467	}}
\author[	85	]{	Ceren	Ulusoy	\orcidlink{	0000-0001-8868-5558	}}
\author[	86	]{	Maio	Umberto			} 
\author[	87	]{	Jochen	Weller	\orcidlink{	0000-0002-8282-2010	}}
\author[	88	]{	Sarod	Yatawatta	\orcidlink{	0000-0001-5619-4017	}}
\author[	89	]{	Shintaro	Yoshiura	\orcidlink{	0000-0003-0581-5973	}}
\author[	90	]{	Bin	Yue	\orcidlink{	0000-0002-7829-1181	}}
\author[7,95]{Saleem Zaroubi\orcidlink{0000-0001-9121-8467}}
\author[	70	]{	Qian Zheng	\orcidlink{	0000-0001-5345-6050	}}

\affiliation[1]{INAF-Istituto di Radioastronomia, via Gobetti 101, 40129 Bologna, Italy}
\affiliation[2]{Centre for Radio Astronomy Techniques and Technologies, Department of Physics and Electronics, Rhodes University, Makhanda, 6140 South Africa}
\affiliation[3]{
South African Radio Astronomy Observatory, Liesbeek House, River Park, Mowbray, Cape Town, 7700, South Africa}
\affiliation[4]{Department of Astronomy, Astrophysics, and Space Engineering, Indian Institute of Technology Indore, 453552, Indore, India}
\affiliation[5]{Institute of Astronomy, University of Cambridge, Madingley Road, Cambridge CB3 0HA, UK}
\affiliation[7]{Kapteyn Astronomical Institute, University of Groningen, PO Box 800, NL-9700 AV Groningen, the Netherlands}
\affiliation[8]{Department of Astronomy, University of California, Berkeley, CA, USA}
\affiliation[9]{Department of Physics and Trottier Space Institute, McGill University, 3600 University Street, Montreal, QC H3A 2T8, Canada}
\affiliation[10]{Department of Astronomy, Tsinghua University, Beijing 100084, People's Republic of China}
\affiliation[11]{Department of Astronomy and Oskar Klein Centre, AlbaNova, Stockholm University, SE-10691 Stockholm, Sweden}
\affiliation[12]{Kapteyn Astronomical Institute, University of Groningen, PO Box 800, NL-9700 AV Groningen, the Netherlands}
\affiliation[13]{LUX, Observatoire de Paris, PSL Research University, CNRS, Sorbonne Universit\'{e}, F-75014 Paris, France}
\affiliation[14]{Department of Physics and Astronomy `Ettore Majorana', University of Catania, Via Santa Sofia 64, 95123  Catania, Italy}
\affiliation[15]{Department of Physics, Blackett Laboratory, Imperial College London, London SW7 2AZ, UK}
\affiliation[16]{Max-Planck-Institut für Radioastronomie, Auf dem Hügel 69, D-53121 Bonn, Germany}
\affiliation[17]{Department of Astrophysics, University of Zurich, Winterthurerstrasse 190, 8057 Zurich, Switzerland}
\affiliation[18]{International Centre for Radio Astronomy Research, Curtin University, Bentley WA, Australia}
\affiliation[19]{CSIRO Space and Astronomy, Kensington WA, Australia}
\affiliation[20]{State Key Laboratory of Radio Astronomy and Technology, National Astronomical Observatories, Chinese Academy of Sciences, A20 Datun Road, Chaoyang District, Beijing 100101, People's Republic of China}
\affiliation[21]{Research Center for the Early Universe, Graduate School of Science, The University of Tokyo, 7-3-1 Hongo, Bunkyo, Tokyo 113-0033, Japan}
\affiliation[22]{Department of Physics, Graduate School of Science, The University of Tokyo, 7-3-1 Hongo, Bunkyo, Tokyo 133-0033, Japan}
\affiliation[	21	]{	Berkeley Center for Cosmological Physics, University of California, Berkeley, CA 94720, United States	}
\affiliation[	22	]{	Max Planck Institute for Astrophysics, Garching, Germany	}
\affiliation[	23	]{	School of Physics and Astronomy, Tel-Aviv University, Tel-Aviv, 69978, Israel	}
\affiliation[	24	]{	School of Physics, University of New South Wales, Sydney, NSW 2052, Australia	}
\affiliation[	26	]{	Institute for Particle Physics \& Astrophysics (ETHZ), Wolfgang-Pauli-Str 27, 8093 Zurich, Switzerland	}
\affiliation[	27	]{	Faculty of Electrical Engineering, Mathematics and Computer Science, Delft University of Technology, Mekelweg 4, 2628 CD Delft, The Netherlands	}
\affiliation[	28	]{	INAF, Istituto di Radioastronomia, Via Piero Gobetti 101, 40129 Bologna, Italy	}
\affiliation[	29	]{	INAF - Osservatorio Astronomico di Trieste, Via G.B. Tiepolo 11, 34131 Trieste, Italy	}
\affiliation[	30	]{	INAF, Istituto di Radioastronomia, Via Piero Gobetti 101, 40129 Bologna, Italy	}
\affiliation[	31	]{	Department of Physics and Trottier Space Institute, McGill University, 3550 rue University, Montr\'{e}al, QC H3A 2A7, Canada	}
\affiliation[	32	]{	Centre for Strings, Gravitation and Cosmology, Department of Physics, Indian Institute of Technology Madras, Chennai 600036, India	}
\affiliation[	33	]{	National Centre for Radio Astrophysics, Tata Institute of Fundamental Research, Ganeshkhind, Pune 411007, India	}
\affiliation[	34	]{	Max Planck Institute for Astrophysics, Karl-Schwarzschild-Str. 1, 85741 Garching, Germany	}
\affiliation[	35	]{	Center for Cosmology and Particle Physics, Department of Physics, New York University, New York, NY 10003, USA	}
\affiliation[	36	]{	Kavli Institute of Cosmology, Institute of Astronomy, University of Cambridge, Madingley Rd, Cambridge, CB3 0HA	}
\affiliation[	37	]{	Department of Physics, Jadavpur University, 188, Raja S.C. Mallick Rd, Kolkata, WB, India	}
\affiliation[	39	]{	Canadian Institute for Theoretical Astrophysics, 60 St George St, University of Toronto, Toronto, ON M5S 3H8, Canada}
\affiliation[	40	]{	David A. Dunlap Department of Astronomy and Astrophysics, University of Toronto, 50 St George St, Toronto ON M5S 3H4, Canada}
\affiliation[	41	]{	Centre for Strings, Gravitation and Cosmology, Department of Physics, Indian Institute of Technology Madras, Chennai 600036, India	}
\affiliation[	42	]{	Institute of Astronomy, University of Cambridge, Madingley Road, Cambridge, CB3 0HA, UK \\ Kavli Institute for Cosmology, Madingley Road, Cambridge, CB3 0HA, UK	}
\affiliation[	43	]{	Department of Astronomy and Oskar Klein Centre, AlbaNova, Stockholm University, SE-10691 Stockholm, Sweden ARCO (Astrophysics Research Center), Department of Natural Sciences, The Open University of Israel, 1 University Road, PO Box 808, Ra’anana 4353701, Israel 	}
\affiliation[	44	]{	Department of Physics, Indian Institute of Technology Kharagpur, Kharagpur 721302, India	}
\affiliation[	45	]{	Department of Astronomy and Oskar Klein Centre, AlbaNova, Stockholm University, SE-10691 Stockholm, Sweden 	}
\affiliation[	46	]{	Universit\'{e} Paris-Saclay, CNRS, Institut d’Astrophysique Spatiale, 91405, Orsay, France	}
\affiliation[	47	]{	Institut f\"ur Theoretische Physik, Universit\"at Heidelberg, Philosophenweg 16, 69120 Heidelberg, Germany	}
\affiliation[	48	]{	Laboratoire de Physique de l’ENS, ENS, Universit\'{e} PSL, CNRS, Sorbonne Universit\'{e}, Universit\'{e}e Paris Cit\'{e}, 75005 Paris, France	}
\affiliation[	49	]{	Department of Astrophysics, University of Vienna, T\"urkenschanzstr. 17, 1180 Vienna, Austria	}
\affiliation[	50	]{	Department of Physics and Astronomy, University of the Western Cape, Robert Sobukwe Road, Bellville 7535, South Africa	}
\affiliation[	51	]{	Guangxi Key Laboratory for Relativistic Astrophysics, School of Physical Science and Technology, Guangxi University, Nanning, Guangxi, 530004, China	}
\affiliation[	52	]{	Department of Astronomy, Astrophysics and Space Engineering, Indian Institute of Technology Indore, Indore - 453552, M.P., India	}
\affiliation[	53	]{	Max-Planck-Institut für Astronomie, Königstuhl 17, 69117 Heidelberg, Germany	}
\affiliation[	54	]{	Department of Astronomy, Astrophysics and Space Engineering, Indian Institute of Technology Indore, Indore - 453552, M.P., India	}
\affiliation[	55	]{	Institute for Astronomy, University of Edinburgh, Royal Observatory, Blackford Hill, Edinburgh EH9 3HJ, UK	}
\affiliation[	56	]{	Max Planck Institute for Astrophysics, Auf dem Hügel 69, 53121 Bonn Germany	}
\affiliation[	57	]{	Department of Physics, Jadavpur University, 188, Raja S.C. Mallick Rd, Kolkata, WB, India	}
\affiliation[	58	]{	Battcock Centre for Experimental Astrophysics, Cavendish Laboratory, J. J. Thomson Avenue, Cambridge CB3 0HE, UK}
\affiliation[	59	]{	Department of Physics, National Institute of Technology Calicut, Calicut 673601, Kerala, India	}
\affiliation[	60	]{	Dipartimento di Fisica e Astronomia, Alma Mater Studiorum Universit\'{a} di Bologna, via Gobetti 93/2, 40129 Bologna, Italy	}
\affiliation[	61	]{	Department of Astronomy, University of Texas at Austin, Texas, USA	}
\affiliation[	62	]{	Research School of Astronomy and Astrophysics, Australian National University, Canberra, ACT 2611, Australia	}
\affiliation[	63	]{	Hangzhou Institute for Advanced Study, UCAS, Hangzhou, 310024, China}
\affiliation[	64	]{	Department of Astronomy, Astrophysics and Space Engineering, Indian Institute of Technology Indore, Indore - 453552, M.P., India	}
\affiliation[65]{University of Chinese Academy of Sciences, Beijing, 100190, China	}
\affiliation[	66	]{	Observatoire Astronomique de Strasbourg, Universit\'{e} de Strasbourg, CNRS UMR 7550, 11 rue de l’Universit\'{e}, 67000 Strasbourg, France	}
\affiliation[	67	]{	Department of Astronomy, Astrophysics and Space Engineering, Indian Institute of Technology Indore, Indore - 453552, M.P., India	}
\affiliation[	68	]{	Institut f\"ur Theoretische Physik, Universit\"at Heidelberg, Philosophenweg 16, 69120 Heidelberg, Germany 	}
\affiliation[	69	]{	Physics Department, Stellenbosch University, 42 Merriman Ave, Stellenbosch, South Africa, 7600	}
\affiliation[	70	]{	Shanghai Astronomical Observatory, Chinese Academy of Sciences	}
\affiliation[	72	]{	Department of Astronomy, Astrophysics and Space Engineering, Indian Institute of Technology Indore, Indore 452020, India	}
\affiliation[	73	]{	Department of Astronomy, National Research Institute of Astronomy and Geophysics (NRIAG), Cairo, Egypt	}
\affiliation[	74	]{	Department of Physics \& Astronomy, University of the Western Cape, Robert Sobukwe Road, Cape Town 7535, South Africa}
\affiliation[	75	]{	LUX, Observatoire de Paris, Universit\'e PSL, Sorbonne Universit\'e, CNRS, 75014 Paris, France	}
\affiliation[	76	]{	Physics and Applied Mathematics Unit, Indian Statistical Institute, 203 B.T. Road, Kolkata 700 108, India	}
\affiliation[	77	]{	School of Space and Environment, Beihang University, Beijing, China 	}
\affiliation[	78	]{	Max Planck Institute for Astrophysics, Karl-Schwarzschild-Str. 1, 85741 Garching, Germany	}
\affiliation[	79	]{	South-Western Institute for Astronomy Research, Key Laboratory of Survey Science of Yunnan Province, Yunnan University, Kunming, Yunnan 650500, Peopleʼs Republic of China	}
\affiliation[	81	]{	INAF–Osservatorio Astronomico di Trieste, Via G.B. Tiepolo, 11, I-34143 Trieste, Italy}
\affiliation[	82	]{	CIERA and Department of Physics and Astronomy, Northwestern University, 1800 Sherman Ave., Evanston, IL 60201, USA	}
\affiliation[	83	]{	Department of Astronomy, Astrophysics and Space Engineering, Indian Institute of Technology Indore, Indore - 453552, M.P., India	}
\affiliation[	84	]{	INAF, Istituto di Radioastronomia, Via Piero Gobetti 101, 40129 Bologna, Italy	}
\affiliation[	85	]{Department of Physics, Eastern Mediterranean University 99628, Famagusta, N. Cyprus}
\affiliation[	86	]{	Italian National Institute of Astrophysics - Observatory of Trieste, via G. Tiepolo 11, 34141 Trieste, Italy	}
\affiliation[	87	]{	Universit\"{a}ts-Sternwarte, Fakult\"{a}t f\"{u}r Physik, Ludwig-Maximilians Universit\"{a}t, Scheinerstraße 1, 81679 M\"{u}nchen, Germany, \\ Max-Planck-Institut f\"ur extraterrestrische Physik, Giessenbachstr.~1, 85748 Garching, Germany	}
\affiliation[	88	]{	ASTRON, Netherlands Institute for Radio Astronomy, Oude Hoogeveensedijk 4, 7991 PD, Dwingeloo, The Netherlands.	}
\affiliation[	89	]{	Institute for Advanced Research, Nagoya University, Furo-cho Chikusa-ku, Nagoya 464-8601, Japan, Graduate School of Science, Division of Particle and Astrophysical Science, Nagoya University, Furocho, Chikusa-ku, Nagoya, Aichi 464-8602, Japan	}
\affiliation[	90	]{	State Key Laboratory of Radio Astronomy and Technology, National Astronomical Observatories, Chinese Academy of Sciences, 20A Datun Road, Chaoyang District, Beijing 100101, China	}
\affiliation[91]{Department of Physics, 60 St George St, University of Toronto, Toronto, ON M5S 3H8, Canada	}
\affiliation[92]{INFN, Sezione di Trieste, Via Valerio 2, I-34127 Trieste, Italy}
\affiliation[93]{SISSA, International School for Advanced Studies, Via Bonomea 265, 34136 Trieste, Italy}
\affiliation[94]{IFPU, Institute for Fundamental Physics of the Universe, Via Beirut 2, I-34151 Trieste, Italy}
\affiliation[95]{Department of Natural Sciences, The Open University of Israel, 1 University Road, Ra’anana 4353701, Israel}

\emailAdd{cathryn.trott@curtin.edu.au}

\abstract{We provide an overview of the eight SKAO Science Book chapters that motivate the Epoch of Reionisation and Cosmic Dawn experiments with SKA-Low. We describe the individual SKA-Low experiments and expected sensitivity - power spectrum, tomography, 21-cm forest, cross-correlations, building on the broad observational plan laid out in the 2015 SKA Science Book. Finally, we outline features of the telescope that will be critical for the success of EoR/CD science, e.g., beam apodization, substations, and multi-beaming.}

\begin{document}
\maketitle

\section{Introduction}

Exploration of the early Universe with the 21~cm line of neutral hydrogen remains a field of research with considerable observational and theoretical effort. As the most abundant element in the Universe, and optically thin across redshift, neutral hydrogen is an excellent tracer of the state of the intergalactic medium (IGM), encoding the thermal and ionisation history of the IGM as a function of space and time, and providing a complementary probe to optical/infrared/sub-mm measurements of galaxies. In the Cosmic Dark Ages ($z > 30$), before the first luminous sources, in the Cosmic Dawn ($z=30-15$), when the first stars and galaxies started to illuminate the cosmos, and during the reionisation phase ($z=15-5.4$), the hydrogen signal is a crucial tracer of the state of the IGM. At early times, the power spectrum of brightness temperature fluctuations traces the matter power spectrum, thereby providing information about cosmology. At late times, during the Cosmic Dawn and Epoch of Reionisation, the power spectrum is dominated by spin-temperature fluctuations and by ionised and neutral regions, which couple astrophysics to the hydrogen hyperfine state. The current generation of low-frequency radio telescopes pursuing this signal does not have the surface brightness sensitivity to fully explore the evolution of the signal over spatial scales and time via direct imaging; instead, they focus on statistical measures that combine data at the expense of information loss, e.g., the spherically-averaged power spectrum of brightness temperature fluctuations.

In this umbrella chapter, we review the key relations that connect observations to astrophysics and present the experimental effort proposed by the Science Working Group to explore the early Universe with the 21~cm line using the SKA-Low AA* and AA4 arrays. Details of the full suite of science enabled by SKA-Low can be found in the other chapters from the EoR/CD Science Working Group. Note that chapter authorship is alphabetical by surname, and contributions of each author are contained within the text:
\begin{enumerate}
    \item \cite{deLeraAcedo2026.SKA}, "Observations of the Cosmic Dawn and Epoch of Reionization with the SKAO: observational lessons from precursors and pathfinder instruments" -- details the considerable observational expertise and knowledge that has been gained over the past 15 years and multiple pathfinder experiments.
    \item \cite{Chakraborty2026.SKA}, "Synergies for the Epoch of Reionization and Cosmic Dawn" -- highlights both the science benefit of complementary probes, as well as how they can be used to mitigate systematics so as to obtain a first 21~cm detection.
    \item \cite{Barkana2026.SKA}, "High-redshift signatures from the Cosmic Dawn and the Epoch of Reionization" -- provides a comprehensive overview of the astrophysical and cosmological processes that shape the 21-cm signal during Cosmic Dawn and the Epoch of Reionization, including both standard and exotic signatures potentially
observable with SKA-Low. 
    \item \cite{Bianco2026.SKA}, "Machine Learning and the SKA for Cosmic Dawn and the Epoch of Reionization" -- outlines advances in machine learning techniques and their application to SKA data and analysis
    \item \cite{Acharya2026.SKA}, "Inferring cosmology and astrophysics from the 21~cm signal with SKA-Low" -- describes the utility of inference techniques for extracting astrophysical information from SKA data
    \item \cite{Cang2026.SKA}, "Exploring the Cosmic Dawn through the 21~cm Forest and high-redshift radio sources with the SKA" -- outlines the key astrophysical information that can be gained from sightlines through the IGM to probe cold gas through absorption against background continuum sources.
    \item \cite{Bag2026.SKA}, "Imaging the 21~cm signal from the Cosmic Dawn and Epoch of Reionization and the connection with the global signal" -- summarizes SKA-Low's abilities to tomographically image the 21-cm signal, the various methods proposed for quantitatively analyzing this data, as well as the how SKA-Low could measure the global, sky-averaged, signal.
    \item \cite{Burba2026.SKA}, "Foregrounds characterization and mitigation in the observations of the CD/EoR with the SKA" -- provides context and learnings from the precursor and pathfinder experiments for foreground understanding and treatment.
\end{enumerate}

 Key contributions to the development of EoR/CD science with SKA-Low, and the experiments to be undertaken, can be found in the first SKA Science Book. These are the foundations upon which we build the experiments today (e.g. \citealt{Ciardi_2015,koopmans15,mellema15}). The \cite{koopmans15} chapter remains relevant for the over-arching experiments, including tiered surveys, balance of beams, spectral bands, and overall observing time for each experiment. Since 2015, however, our theoretical understanding has evolved significantly, and with it our expectations around what is required to explore the signal.  Advances in data science and machine learning are also promising avenues for reducing unknown residuals \citep{lofarres,mwares,herares}. In addition, cross-correlation studies have emerged as fundamental for deeper astrophysical understanding as well as for mitigating the systematics currently limiting the first generation interferometers and thus paving the way for a first 21~cm detection. In tandem with the Book chapters highlighted above, this updated chapter then provides the opportunity to document some of the observational and theoretical considerations that have emerged since that time, now that an array design is in place.

\section{Astrophysics from the 21~cm line}

The high-redshift 21~cm line (i.e. the cosmic 21cm signal), is sourced by the neutral hydrogen that fills the intergalactic medium. As the most abundant element in nature, the evolution of hydrogen reflects the evolution of the Universe. The 21-cm line corresponds to the hyperfine transition in the ground state of neutral hydrogen, resulting from the energy difference between the parallel and anti-parallel spin states of the proton and electron.  
   Its detectability depends on the spin temperature of hydrogen, which quantifies the relative occupancy of the two hyperfine states.  The spin temperature is determined by the properties of the gas and cosmic UV radiation fields, and can be expressed as,
(e.g. \citealt{Field58, 2006PhR...433..181F}):
\begin{equation}
\label{eq:spinT}
    T_S^{-1} = \frac{T_\gamma^{-1} + x_cT_K^{-1} + x_\alpha T_c^{-1}}{1+x_c +x_\alpha},
\end{equation}
where $T_K$ is the gas kinetic temperature and $T_c \sim T_K$ is the colour temperature of the Ly-$\alpha$ radiative field. The coefficients $x_c$ and $x_\alpha$ quantify the coupling of $T_S$ to the temperature of the gas via collisions (only important in the IGM at $z \gtrsim 30$) and absorption and re-emission of Ly-$\alpha$ photons, respectively \citep{Wouthuysen52, Field58}. 
%
 Intensity mapping observations then measure the fluctuations in the brightness temperature, $T_B$, which can be expressed in terms of the spin temperature from eq. (\ref{eq:spinT}), and the radio background temperature, $T_\gamma$  \citep[typically taken to be the CMB, though see Chapter][for more exotic scenarios]{Barkana2026.SKA}.  This brightness temperature contrast against the background is a function of both position and redshift, and can be written as (e.g. \citealt{Furlanetto2006}):
\begin{equation}
\label{eq:deltaT}
    \Delta{T}_B \simeq \frac{T_S - T_\gamma}{1+z}\tau_\nu \simeq 9 x_{\rm HI}(1+\delta)(1+z)^{1/2}\left[ 1-\frac{T_\gamma}{T_S} \right] \left[ \frac{H(z)/(1+z)}{dv_\parallel/dr_\parallel} \right] \text{mK},
\end{equation}
where the final approximation assumes $\tau_\nu \ll 1$, and a cross-section for interaction of negligible width.  Here, $x_{\rm HI}$ is the hydrogen neutral fraction, $(1+\delta) = \rho/\bar{\rho}$ the density contrast, and ${dv_\parallel/dr_\parallel}$ the velocity gradient along the line of sight.  As seen from the above equations, the 21 cm signal is sensitive to astrophysical and cosmological processes that heat and ionize the IGM.  In the most plausible scenario, this is done by X-ray and UV radiation from galaxies.  Because these cosmic radiation fields are sourced by the combined contribution of {\it all} galaxies, the 21-cm signal offers a powerful way of studying not just the IGM directly, but also indirectly the birth and evolution of the first galaxies, the vast majority of which are too faint to be seen even with powerful UV/IR telescopes such as the James Webb Space Telescope (JWST; e.g. \citealt{OShea15, Qin20}).

Our understanding of these galaxies and how they shape the thermal and ionization state of the IGM has evolved dramatically over the past few years.  A wealth of data, most notably from the Lyman alpha forest \citep[e.g.,][]{2022MNRAS.514...55B, Qin2024}, helped us narrow down the timing of reionization, which ends later and begins earlier than previously believed (though see \citealt{Lidz07, Mesinger10}).  JWST has been finding a surprising number of bright galaxies at very high redshifts ($z\geq10$; e.g. \citealt{Donnan23, Leethochawalit23, McLeod24}).  The most popular explanations include increased star formation efficiencies and / or bursty star formation (e.g.  \citealt{Dekel23, FPD23, MTT23, Hutter25}]), with unclear implications for the 21~cm signal (though see \citealt{DMM25, CC25, Dhandha25}).  The upper limits on the 21~cm power spectrum from SKA precursor instruments point to a new source of IGM heating at $z\geq10$ \citep{HERA23, LOFAR25, MWA25}; if this heating is driven by high mass X-ray binaries (HMXBs), as seems most likely, the data would confirm theoretical claims that HMXBs forming in metal poor (i.e., pristine) gas are significantly more X-ray luminous than any found in the local universe (e.g. \citealt{Fragos12}).

The fluctuations in eq. (\ref{eq:deltaT}) can be measured directly (tomography) or statistically, for example through a spatial power spectrum.  Initial measurements are noise dominated, and so the power spectrum metric is commonly used to improve the signal to noise, as well as to separate the cosmic signal from the foregrounds that theoretically should occupy a well-defined region in the 2D power spectrum.  More mature measurements with SKA will result in images, which will fully unlock the potential of this non-Gaussian signal in understanding the early Universe. The power spectrum estimator is the Fourier-dual to the two-point correlation function, and is given by:
\begin{equation}
    P_T(|{\bf k}|) = \frac{1}{\Omega}\langle \tilde{T}_B({\bf k}^\prime) \tilde{T}^\dagger_B({\bf k}) \delta({\bf k},{\bf k}^\prime)\rangle,
\end{equation}
where the ensemble average is over different orientations of the mode of length $|{\bf k}|$, and $\Omega$ is the observing volume. This equation demonstrates how the observing volume and spatial resolution affect the number of modes available for power spectrum estimation, and therefore both the thermal noise and sample variance.

In addition to the intensity mapping mode of observation using the CMB as the background, we also expect to observe 21~cm absorption lines against high-redshift radio point sources during reionization with SKA-Low \citep{Ciardi2015b}. These absorption lines, known as the 21~cm forest in analogy to the Ly-$\alpha$ forest, provide a sensitive probe to the thermal history and small-scale structures in the early Universe (e.g. \citealt{Furlanetto2002,Xu2009,Xu2011MN}). Absorption of the continuum signal by intervening neutral gas is deepest for cold gas (small $T_{\rm K}$) 
  with a high column density ($n_{\rm HI}$). The absorption level is quantified by the optical depth, $\tau$, with the normalised measured signal (transmission) quantified as \citep{Soltinsky_2025}:
\begin{equation}
    F(\nu) = \frac{S_{\rm meas.}(\nu)}{S_0} = \exp{(-\tau)},
\end{equation}
where the optical depth is given by:
\begin{equation}
    \tau = \frac{3h_p c^3A_{10}}{32\pi^{3/2}\nu_{21}^2k_B} \frac{\delta{v}}{H(z)} \displaystyle\sum_{j=1}^N \frac{n_{HI,j}}{b_j T_{S,j}}\exp{-\frac{(v_{H,i}-u_j)^2}{b_j^2}}.
\end{equation}
Because the absorption is relative to the background source flux density, $S_0$, the normalised signal is invariant for given IGM conditions along the line-of-sight. Deviations from a perfect continuum spectrum can then be quantified by the statistics of the mean-subtracted transmission:
\begin{equation}
    \delta_F = \exp{(-\tau)} - 1.
\end{equation}

The line-of-sight (spectral) Fourier transform of this normalised transmission provides information on the spatial scales of the absorbing gas, and we compute the 1D power spectrum:
\begin{equation}
    k_qP(k_q) = k_q\frac{2\pi}{n\Delta\nu} |\tilde\delta_F |^2,
\end{equation}
with $k_q = 2\pi q/n\Delta\nu$ MHz$^{-1}$, $q$ labelling wavemode, and $n$ frequency channels of spectral resolution $\Delta\nu$. The equations presented here are appropriate for data that contain one continuum source with absorption, and radiometric noise only. In practice, the field will contain many sources that contribute spectral sidelobes, yielding a systematic error equivalent to the "EoR wedge", which is a region of wavemode parameter space that is contaminated by smooth spectrum foregrounds \citep{2020ApJ...899...16T}.

Identification of high-redshift radio-loud candidate sources is improving, with radio follow-up of optically identified sources and SKA precursor telescopes playing a role in radio detections \citep{2018MNRAS.480.2733S,2015ApJ...804..118B}. For SKA-Low AA2 and beyond, there will be sufficient candidates to undertake early test observations and refine the observational strategy.

\section{Experiments with SKA-Low AA2, AA* and AA4}

The AA* and AA4 array assemblies offer similar performance for EoR/CD science. While the AA2 array assembly, with its lack of short baselines, is suited to constructing a deep and high-resolution sky model for instrument calibration, which requires good point source sensitivity and low sidelobe contamination, and science verification and testing of data and pipelines. Table \ref{table:experiments} describes the expected types of key experiments that will serve the bulk of the EoR/CD science cases described in the chapters of this book. Statistical measures are expected to be estimated across the full redshift range, while tomography will be available below $z=10$. Frequencies $>220$~MHz will not be used, allowing for extra bandwidth to be traded for extra beams. The visibility datasets are proposed to be calibrated (direction-independent calibration), and lightly averaged, with flagging and calibration sky model metadata attached. Direction dependent calibration is proposed to occur outside of the Observatory, by the science team.
\begin{table}
    \centering
    \begin{tabular}{|c|c|c|c|c|}
    \hline
       Experiment & Redshift & Freq. (MHz) & Station size (m) & Data product \\\hline
       $T_B$ power, poly spectra  & $5.5 <z<27$ & 50--220 & 38, 15--18 & Visibilities, 50~kHz, 2s\\
       21cm Forest  & $5.5<z<10$ & 130--220 & 38 & Image cubes, 5~kHz\\
       HI tomography  & $5.5<z<15$ & 90--220 & 38, 15--18 & Visibilities, 50~kHz, 2s\\
       \hline
    \end{tabular}
    \caption{Expected suite of experiments that will serve most of the science presented in EoR/CD chapters of this book. Statistical measures are expected to be estimated across the full redshift range, while tomography will be available below $z=10$. Frequencies $>220$~MHz will not be used, allowing for extra bandwidth to be traded for extra beams. `Visibilities' refers to DI-calibrated (direction independent) and lightly averaged visibilities.}
    \label{table:experiments}
\end{table}

The interpretation of these visibility, image, and power spectrum datasets is discussed in the Inference \citep{Acharya2026.SKA} and Machine Learning \citep{Bianco2026.SKA} chapters. The Inference chapter reviews state-of-the-art simulation frameworks---including analytical, semi-numerical, numerical, and emulator-based approaches---alongside inference techniques for a range of statistical probes, from the power spectrum to higher-order statistics and field-level analyses. The Machine Learning chapter surveys applications across the full data processing pipeline, from calibration, foreground removal, and RFI mitigation to simulation-based inference and emulation.

Both visibility- and image-based datasets will also be used for cross-correlation of 21~cm with other tracers, as described in the Synergies chapter \cite{Chakraborty2026.SKA} and summarised here. Radio data can be correlated with cosmic backgrounds (e.g., CMB, Near Infrared), resolved galaxy surveys, line intensity mapping data (e.g., star formation tracers such as CO, CII), and quasar spectra (e.g., for 21~cm Forest studies).
Scientifically, there is therefore a rich suite of experiments that can provide astrophysical information on a range of scales. Cross-correlation of 21~cm data also has several practical advantages, including providing independent systematics that may break key degeneracies and matching radio data with a known signal to strengthen confidence in a detection.
Key challenges of cross-correlation include mismatched fields of view, sampling functions, and spatial and spectral resolutions of datasets, which require careful data treatment and analysis to avoid incorrect conclusions \citep{2024ApJ...975..222F}. For the SKA-Low, which is likely to begin operations before there is a significant number of 21~cm power spectrum detections at $z>6$, cross-correlation can play a key early role in guiding observations and confirming (first) detections.

\subsection{Data requirements}

The EoR/CD community requires visibility measurements for 21~cm science. Visibilities are the natural measurement space of an interferometer, sampling the Fourier transform of angular scales as a function of frequency, thereby encoding lookback time. These raw data product dimensions are $(u, v, (w), f)$, where the curvature Fourier dimension $w$ will be handled through the measurement equation. The data products for 21~cm science with SKA-Low are drawn from these frequency-dependent visibilities:
\begin{itemize}
    \item Power spectrum (and other higher-order statistics): $(u,v,f) \rightarrow (u,v,\eta) \propto (k_x, k_y, k_z)$;
    \item 21cm Forest (spectrum): $(u,v,f) \rightarrow (l, m, f)$;
    \item 21~cm Tomography: $(u,v,f) \rightarrow (l, m, z)$;
    \item 21~cm/cross-tracer cross-correlation: $(u,v,f) \rightarrow (l, m, z)$,
\end{itemize}
where $l, m$ are the direction cosines in the sky frame.

\begin{figure}[t!]
    \centering
    \includegraphics[scale = 0.45]{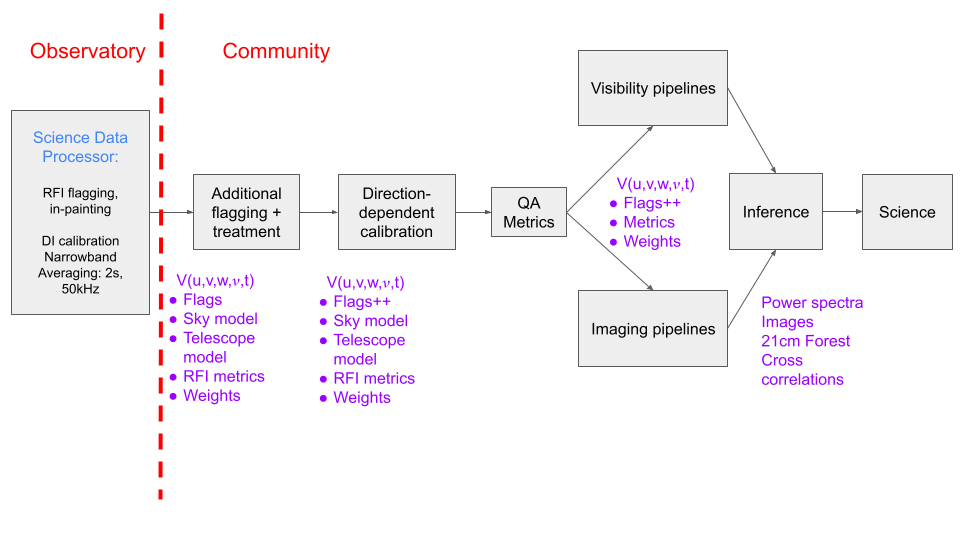}
    \caption{Flowchart of data and pipelines envisaged for SKA-Low experiments. Data from the SKAO Science Data Processor are expected to be DI-calibrated and averaged visibilities. The direction-dependent treatment of the data, quality assurance metrics, data products, and interpretation and inference will occur outside of the Observatory, as part of the SRC Network and community-based pipelines.}
    \label{fig:flowchart}
\end{figure}

Figure \ref{fig:flowchart} shows a schematic of the data flow for EoR/CD experiments. The direction-dependent treatment of the data, quality assurance metrics, data products, and interpretation and inference will occur outside of the Observatory, as part of the SRC Network and community-based pipelines.
From an instrumentation perspective, measurements of extended low-surface-brightness structures (power spectrum and tomography) demand short baselines and excellent surface-brightness sensitivity over scales ranging from tens of arcseconds to degrees, a design requirement that drives much of the SKA-Low core layout. The 21~cm forest experiment, however, is a point source experiment, and will require some longer baselines for precise sightlines (although very long baselines introduce spectral sidelobes of other sources in the sky (i.e., mode-mixing), potentially leading to systematics domination of the spectrum, \cite{2020ApJ...899...16T}). Further details are provided in the chapter by \cite{Cang2026.SKA}.

\subsection{Sensitivity of SKA-Low for high-redshift 21~cm experiments}

\begin{figure}[t!]
    \centering
    \includegraphics[scale = 0.3]{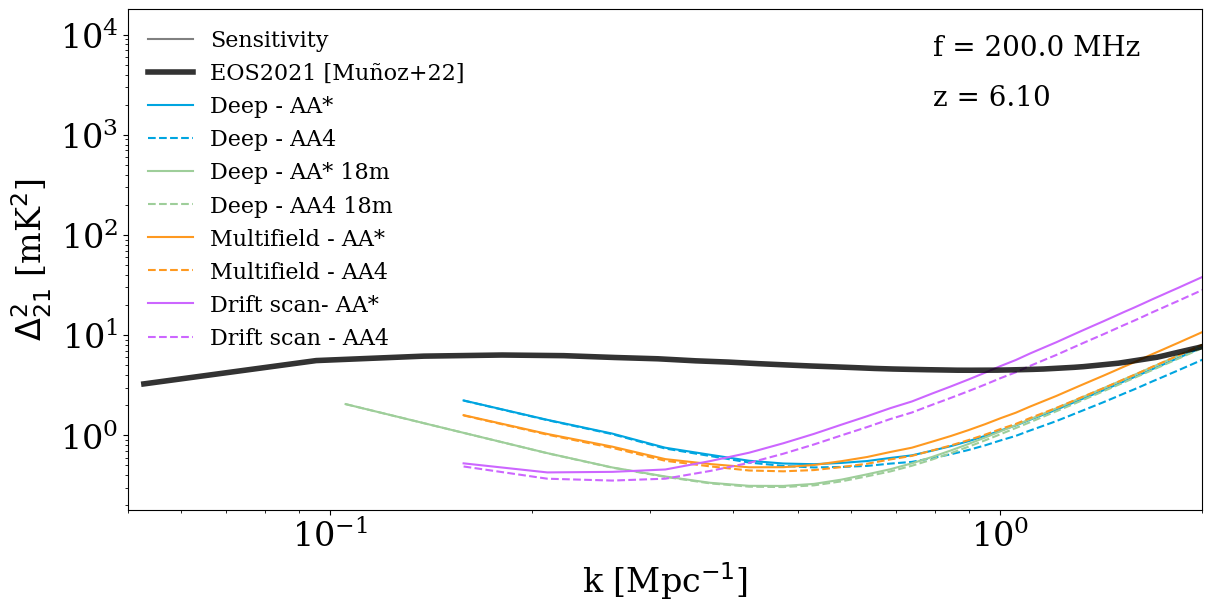}
    \caption{Sensitivity curves for 1000~h experiments with SKA-Low AA* and AA4 at $z=6.1$, as a function of spatial wavenumber. These sensitivity estimates assume a foreground avoidance strategy, cutting out all modes below the horizon plus a 0.1$h$/Mpc buffer. Experiments with 18~m substations (green) allow for measurements at larger angular scales. The black curve shows a simulated 21~cm signal from the EOS2021 simulation of \citealt{Munoz22}.}
    \label{fig:sens_sub}
\end{figure}

\begin{figure}[t!]
    \centering
    \includegraphics[scale = 0.3]{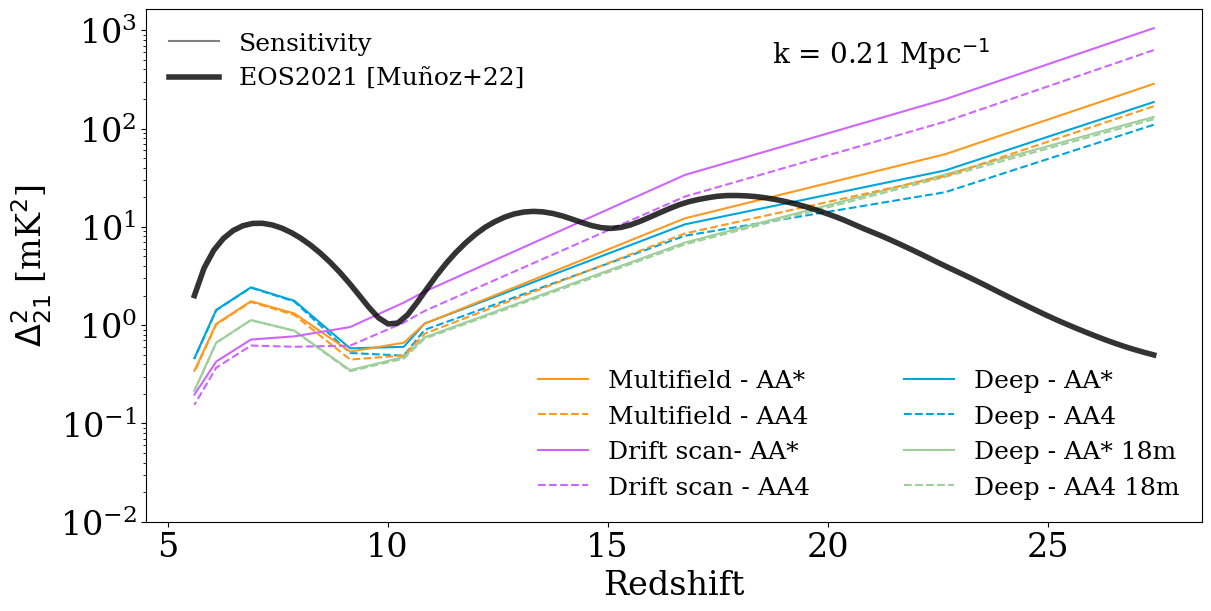}
    \caption{Sensitivity curves for 1000~h experiments with SKA-Low AA* and AA4 at $k=0.21 \textrm{Mpc}^{-1}$, as a function of redshift. Colours are the same as in Figure \ref{fig:sens_sub}.}
    \label{fig:sensitivity}
\end{figure}

SKA-Low has the design and sensitivity to fully explore the EoR and Cosmic Dawn eras, with both statistical and direct-detection experiments of the redshifted 21-cm signal. Its performance relative to the current generation of instruments is described in the chapter by \cite{deLeraAcedo2026.SKA}. There are many calculators that can predict sensitivity performance for each array assembly, including both radiometric noise and sample variance. Example sensitivity curves for different experiments and array assemblies for the power spectrum experiments are shown in Figures \ref{fig:sens_sub} and \ref{fig:sensitivity}, as a function of spatial wavemode and redshift, respectively, using the SKA module of the 21cmSense tool\footnote{https://github.com/rasg-affiliates/21cmSense} \citep{2016ascl.soft09013P,2024JOSS....9.6501M}. These calculators make modest assumptions about loss of parameter space due to foregrounds, but experience with precursors has shown that these estimates are optimistic, and actual performance will be degraded. Nonetheless, SKA-Low's superior sensitivity and flexibility enable experiments and analyses to be refined to deliver transformational science.

\section{Key features of AA* and AA4 for EoR science}

SKA-Low will be the most flexible low-frequency telescope to date. It is the combination of its excellent snapshot and integrated surface-brightness sensitivity on scales from arcminutes to degrees, its broadband frequency coverage, and its flexibility to shape and resize the sky response function that makes SKA-Low an ideal telescope for EoR/CD experiments. Other crucial features that EoR/CD experiments would use include instrumental spectral smoothness \citep{barry16,eloy17,trottwayth16}, channel-based calibratability, and non-chromatic instrumental sidelobes. The latter corresponds to the frequency dependence of the sidelobes of the telescope's primary beam, at angles from the zenith where sky-based calibration is inaccurate. Sidelobe spectral structure imparts spectral chromaticity to residual signals outside the main calibration field, mimicking the 21~cm signal and leaking power into the “EoR Window”, an area of parameter space where the cosmological signal can be separated from smooth-spectrum continuum sources (see, e.g., \citealt{2020PASP..132f2001L} and references therin). SKA-Low has the ability to "apodize" a station response to improve chromaticity of sidelobes, however this is imposed at the expense of sensitivity, thereby requiring longer observing time for the same signal-to-thermal-noise ratio.

\begin{figure}[t!]
    \centering
    \includegraphics[scale = 0.3]{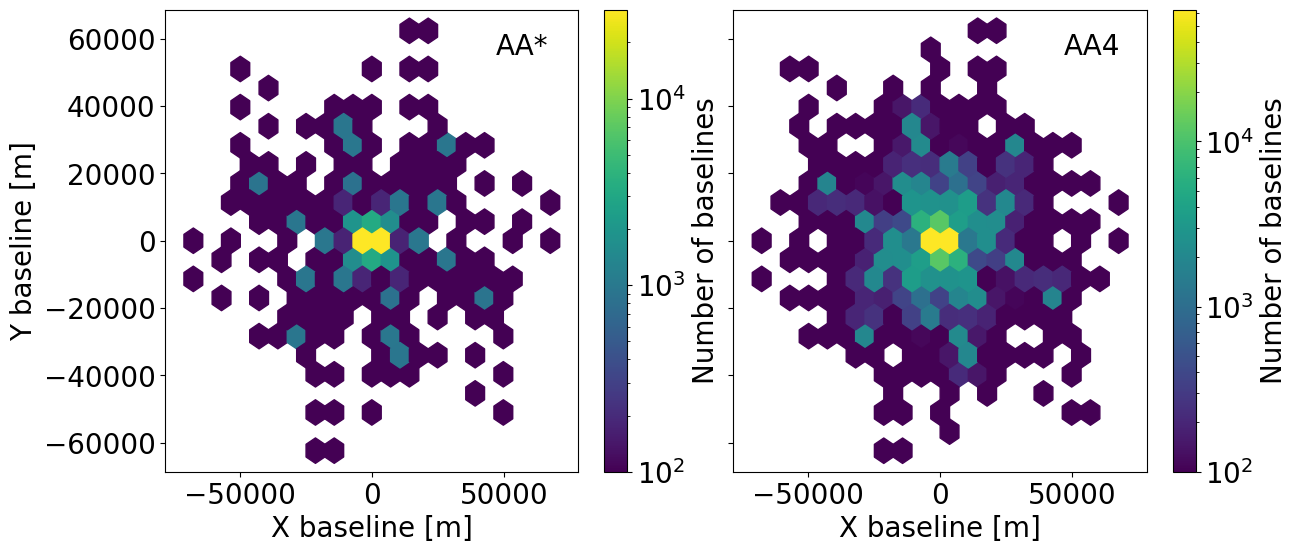}
    \caption{Baseline distribution for a zenith snapshot with AA* (left) and AA4 (right). From D. Breitman.}
    \label{fig:layout}
\end{figure}

The AA* and AA4 arrays yield similar estimation performance for EoR/CD tasks (see Figure \ref{fig:sensitivity}). AA4 affords some incremental core sensitivity, but does not significantly alter the uv-coverage or present other qualitatively different features. The principal features of AA*/AA4 for EoR/CD science are (i) long baselines to build a deep and complete sky model for instrument calibration, and point source sensitivity for 21~cm forest experiments; (ii) filled aperture coverage in the core (central 1~km), yielding excellent snapshot surface brightness sensitivity on scales of relevance for intergalactic neutral gas. AA* has poorer surface-brightness sensitivity than AA4, which more closely resembles the filled-aperture (sea-of-elements) concept envisaged for the SKA-Low core. Figure \ref{fig:layout} shows the baseline-density of the core of the telescope for the AA* and AA4 configuration\footnote{\url{https://21cmsense.readthedocs.io/en/latest/tutorials/SKA_forecast.html}}.
In addition to the raw sensitivity of the base arrays, there are other aspects of the system that EoR/CD will use \citep{trott_2024_16951143}:
\begin{itemize}
    \item Substations - the ability to split a 35~m station of 256 dipoles into smaller units, consequently leading to an increase in field-of-view and the availability of shorter baselines, but generally leading to a lower sensitivity due to loss in total collecting area;
    \item Multi-beaming - the ability to have more than one beam on the sky, at the cost of reduced bandwidth or number of stations;
    \item Beam apodization - the ability to apply frequency-dependent complex weights to the dipoles in a station to shape the frequency-dependent primary beam, also increasing the field of view, but decreasing the station's sensitivity;
    \item Drift scans and pointed observations - flexibility to trade-off thermal noise sensitivity and sample variance.
\end{itemize}

\subsection{Substations}

\begin{figure}[t!]
    \centering
    \includegraphics[scale = 0.35]{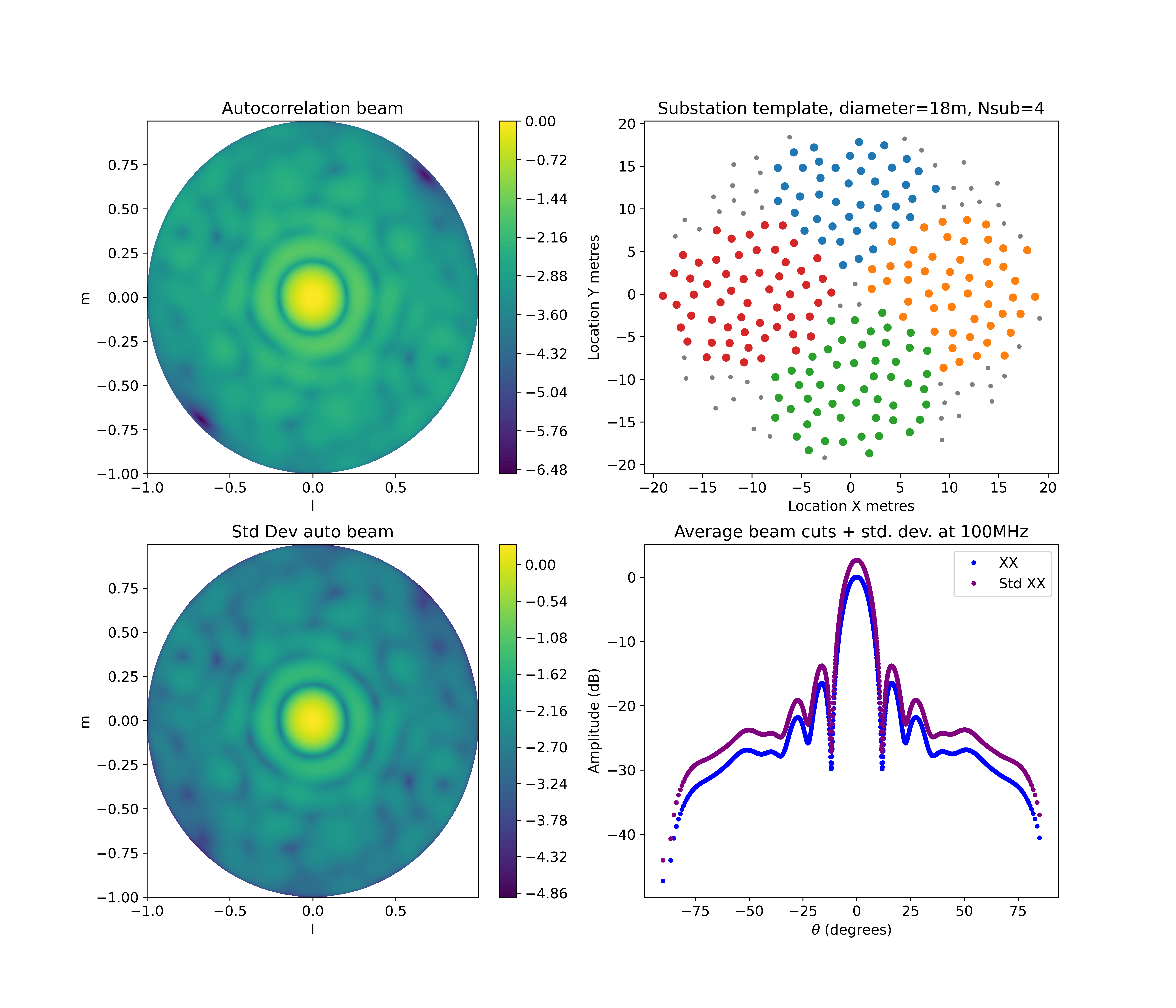}
    \caption{Potential substation configuration for science at the end of reionisation. Here, 18m diameter substations are composed from a full station, with the increased primary beam response shown in the top-right, and its variability across the four substations in the bottom-left.}
    \label{fig:substations}
\end{figure}

Each SKA-Low dipole is digitised, allowing a beamformed ``station'' to be flexibly defined across different numbers of dipoles. Reducing the station size with fewer dipoles yields larger fields of view and shorter baselines. Near the end of reionisation, the characteristic sizes of structures are expected to have a comparable angular size to the primary beam at 190--210~MHz \citep{2023ApJ...954L..14H,2025arXiv251018946N}. As such, forming larger fields of view with shorter baselines provides access to scales of scientific interest that would otherwise be spatially inaccessible. At some redshifts, EoR/CD SWG recommends to divide each full station in the core into a small number of substations (3--4), with 15--18~m diameters, as presented in the SKAO Subarray and Substation documentation. Figure \ref{fig:substations} shows an example of the 18~m substation configuration and beam at 100~MHz.

\subsection{Multi-beaming}

The proposed EoR/CD experiments would span 50-220 MHz for their science and provide a sufficient spectral lever arm for smooth spectrum calibration. For a particular experiment, $\sim$10-15~MHz of instantaneous bandwidth can be used for scientific analyses, avoiding excessive evolution over the cosmic time spanned by the low- and high-frequency limits of the band, and $\sim$50–100~MHz can be used for calibration and foreground characterisation. As such, EoR/CD would form 2–4 beams on the sky simultaneously while retaining sufficient bandwidth for calibration and science. SKA-Low has 300~MHz of instantaneous bandwidth, but EoR/CD SWG would use only 75--150~MHz at any one time, enabling multiple beams to formed on the sky "for free". This would allow for analysis of 2-4 separate fields. With at least $\sim$1000-hour observations required on a single deep field for Cosmic Dawn science and EoR tomography in AA4, the ability to trade bandwidth for incoherent observations over multiple fields allows a trade-off of radiometric noise, redshift coverage and sample variance \citep[e.g., see][]{koopmans15}.

\subsection{Beam apodization}

\begin{figure}[h!]
    \centering
    \includegraphics[scale = 0.25]{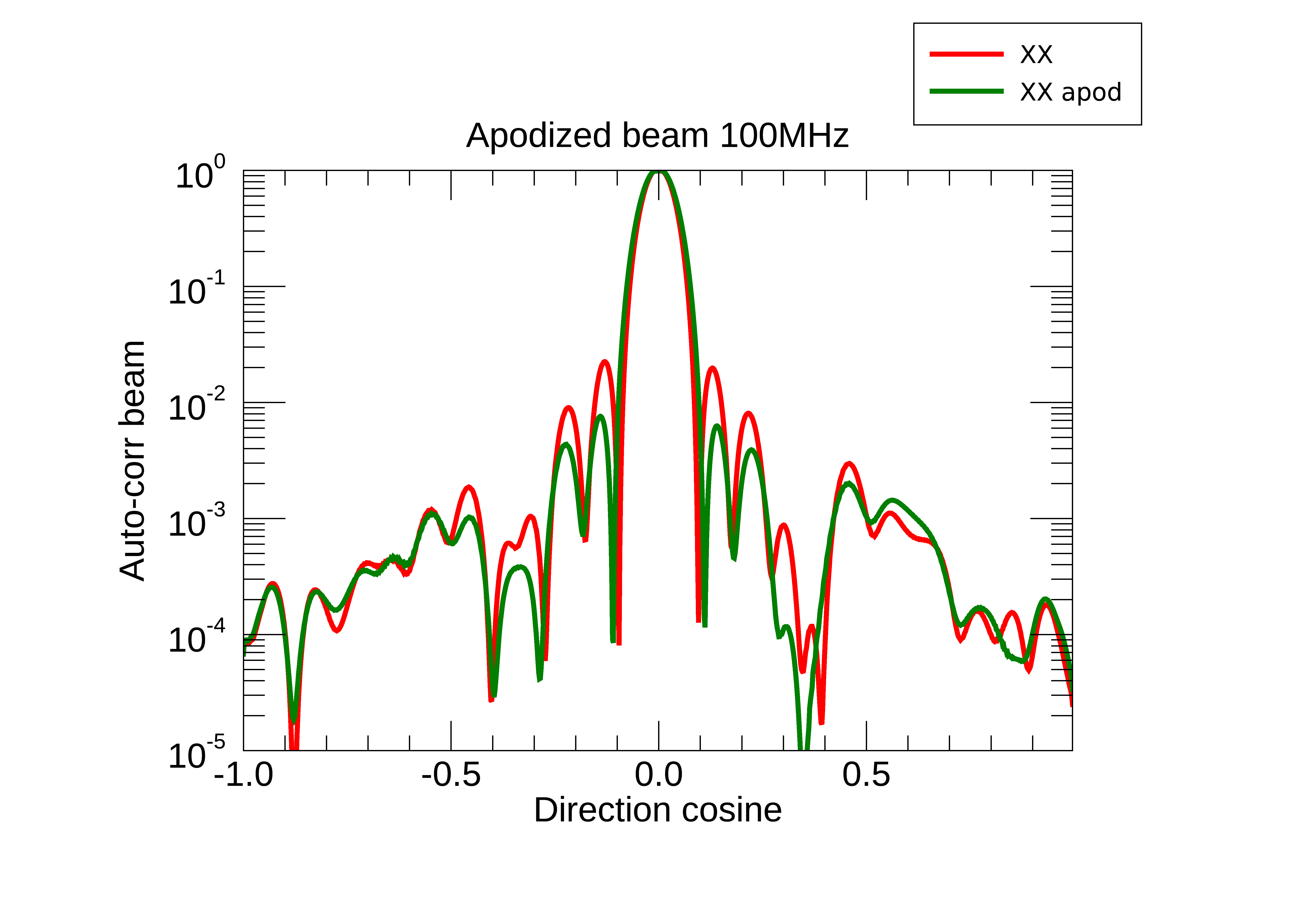}
    \includegraphics[scale = 0.25]{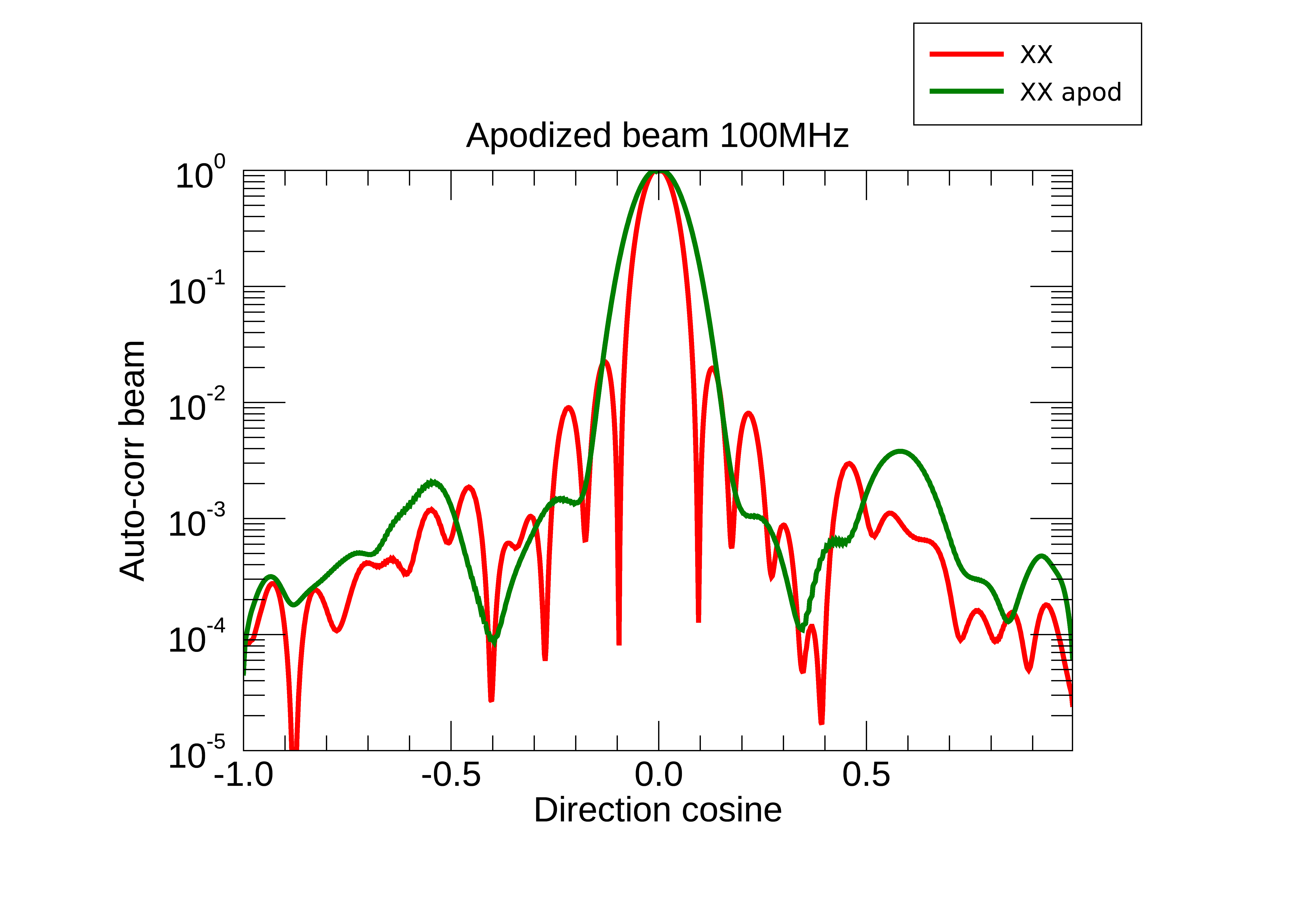}
    \caption{Examples of station beam apodization at 100~MHz and the East-West polarisation. (Left) Real-valued Gaussian taper with characteristic size of the station diameter; (right) real-valued Gaussian taper with characteristic size of half of the station diameter. In general, apodization smooths the beam, reduces inner sidelobes, but does not change outer sidelobes, which are controlled by the discretisation of the dipoles.}
    \label{fig:apodization}
\end{figure}

Digitisation of dipoles allows individual complex beam weights to be applied to each dipole (apodization), enabling subsequent shaping of the primary beam, at the potential cost of sensitivity. Given that the EoR/CD experiments require spectral smoothness and low residual sidelobe chromaticity, this feature is useful. \textit{Frequency-dependent complex beam weights provide the most flexibility for the observer to shape the beam for their science, including the ability to produce a spectrally achromatic beam in some parts of the sky.}
Examples of station beam apodization at 100~MHz and the East-West polarisation are shown in Figure \ref{fig:apodization}, including (left) a real-valued Gaussian taper with a characteristic size of the station diameter, and (right) a real-valued Gaussian taper with a characteristic size of half the station diameter. In general, apodization smooths the beam, reduces inner sidelobes, but does not change outer sidelobes, which are controlled by the discretisation of the dipoles.

\subsection{Drift and tracked scans}


The proposed EoR/CD experiments would also leverage both tracked and drift scans. In the former, the phase centre of the observation (a point on the celestial sphere) is tracked, and the beam-pointing centre tracks it as well. Such an observation mimics a typical dish-based observation. In the latter, drift scans will typically set the pointing centre to the zenith, with the sky rotating overhead. The phase centre can be set to the zenith or phased to track across a small angular range within the telescope's primary beam to retain some short coherence. The tracked observations afford the most coherent averaging of the data, with a consequent reduction in the noise level, because the phase centre is unchanged, but at the cost of a time-dependent beam shape. The drift-scan observations cannot coherently track a single field over long timescales (20 minutes for 5 degrees), but they maintain a more stable single-beam model throughout the observation, which may be more straightforward to handle in the analysis. However, it trades-off increased sky coverage against integration time per unit area.

\section{Outlook}

The EoR/CD Science Working Group has had a coherent vision for the experiments and science to be undertaken with SKA-Low since 2014. With new results and learnings from precursor telescopes and experiments, such as LOFAR, MWA, HERA, and NenuFAR, we have refined the observational strategy and added cross-correlation studies as one of the main science programs. SKA-Low is the only telescope capable of imaging the 21~cm brightness temperature distribution and has the flexibility and modularity needed to shape the observations to maximise science and minimise systematics. Recent advancements with other probes of reionisation and cosmic dawn have engendered tantalizing scientific puzzles that have demonstrated considerable interest within the broader astrophysics community; SKA-Low will only further stimulate activity in this area with a complementary view of the cosmos, ushering in an irrevocable change in our picture of the high-redshift universe in the next decade.

\bibliographystyle{abbrvnat}
\bibliography{chapter} 

\appendix

\end{document}